# Molybdenum low resistance thin film resistors for cryogenic devices

Yu P Korneeva[1,*], M A Dryazgov[1], N V Porokhov[1,2], N N Osipov[1], M I Krasilnikov[1], A A Korneev[1,3], M. A. Tarkhov[1,2]

[1] Institute of Nanotechnology of Microelectronics of the Russian Academy of Sciences, Nagatinskaya 16A, build. 11, Moscow, 115487, Russia
[2] National Research University MPEI, Krasnokazarmennaya 14, build. 1, Moscow, 111250, Russia
[3] Higher School of Economics - National Research University, Myasnitskaya 20, Moscow, 101000, Russia

E-mail: korneeva_yuliya@mail.ru

**Abstract**

We present a study of thin-film molybdenum resistors for NbN electronics operating at cryogenic temperatures. The key step is the 0.5-1.5 keV ion cleaning-activation of NbN before Mo deposition which allows to obtain a high-quality Mo/NbN interface which together with additional aluminum bandage layer in the area of contact pads allow to reduce contact resistance below 1 Ω. The quality of the interfaces is confeirmed by transmission electron microscopy and X-ray reflectometry.

Keywords: thin-film resistors, superconducting films, contact resistance

## 1. Introduction

Cryogenic thin-film resistors are components of various superconducting integrated circuits and devices. In RSFQ (Rapid Single Flux Quantum) circuits and superconducting quantum processors the resistors are used for shunting Josephson junctions and for fabrication of RC filters [1]. In superconducting single photon detectors [2, 3], resistors provide impedance matching required for the recovery of superconductivity in the detector with low kinetic inductance after photon absorption. Many applications require resistors as small as a few Ohms. For reproducible fabrication of such resistors, the resistance of the superconductor/normal metal interface is critical [4, 5].

Two categories of materials have been used for the fabrication of cryogenic thin film resistors: (1) metals such as Mo, Ta, Pd and (2) metal alloys such as NiCr, MoNx, AuPd. Although metal film fabrication is a well-developed and mature technology, the reliable fabrication of thin film resistors operating in wide temperature range including cryogenic temperatures is still an important challenge. The reactive magnetron sputtering is usually used as it enables the control over the film resistance by the adjustment of sputtering parameters.

We present our study of the fabrication technology of molybdenum resistors for NbN superconducting devices. We investigate the influence of fabrication process on the quality of the interfaces between the resistor and the superconducting NbN film. For our research, the molybdenum is chosen because of good process reproducibility and low resistivity (20 - 25 μΩ·cm) for films with thickness of a few tens of nanometers (20-80 nm). Molybdenum has proven itself as a material for planar resistors in Nb/Al-Al-AlO$_x$/Nb Josephson junctions. It is used by such commercial companies as Northrop Grumman Space Technology (NGST) [6], HYPRES Inc. [7, 8], The European FLUXONICS Foundry and by many research groups, e.g. MIT Lincoln Laboratory [9], National Institute of Advanced Industrial Science and Technology in Japan [10], etc.

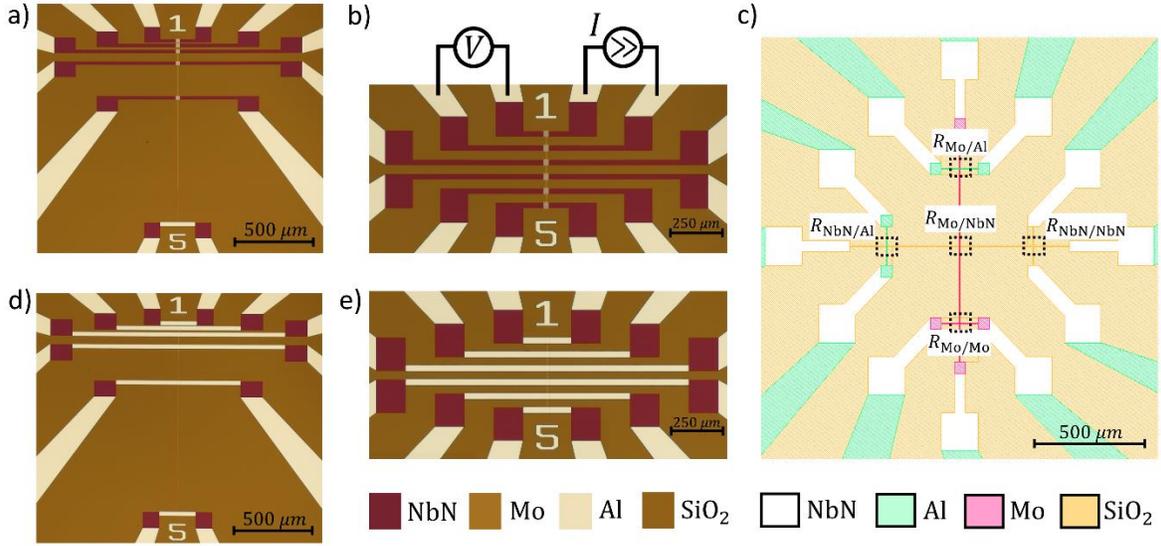

**Figure 1.** Optical photographs of the samples under study: (1) five resistors 3 µm wide, length from 3 to 768 µm (a) without an aluminum bandage and (d) with an aluminum bandage. (2) Five resistors of the same size (b) without an aluminum bandage and (e) with an aluminum bandage. Panel (b) also shows a schematic of a circuit for four-probe resistance measurement (labels "*I*" and "*V*" indicate current and potential contacts). (3) Test structure for measuring the resistance of various interfaces (c), the location of the tested interfaces is shown in the photo by dashed rectangles.

## 2. Fabrication process

We fabricated three types of structures: (1) resistors of fixed width and varying lengths, which is the traditional topology to determine the sheet resistivity $R_s$ of the film and the interface contact resistance $R_0$ (Fig. 1(a) and (d)); (2) five resistors of equal size to determine the reproducibility of the proposed fabrication process (Fig 1(b) and (e)); (3) perpendicularly intersecting strips of different metals in combinations NbN/Al, NbN/Mo, Mo/Al to test the resistance of each interface separately (Fig 1 (c)). The width of the resistors of type (1) and (2) is 3 µm with the contact pads area 20 µm × 26 µm. In type (3) structures intersecting areas vary from 3 to 30 µm, the size of the contact pads is 30 µm × 30 µm.

Each type of resistors is fabricated in two versions: with direct contact of Mo to the NbN film (Fig 1 (a) and (b)) and with an additional layer of aluminum deposited on top of the Mo/NbN contact area to reduce the contact resistance (Fig 1 (d) and (e)). In literature, this technique is called "bandage", it is used in the fabrication of contacts to Josephson junctions [11] to eliminate a lossy interface layer.

The samples under study are prepared on a low-resistivity silicon substrate, pre-oxidized in oxygen plasma to 250 nm thick $SiO_2$ layer at a temperature of 1000°C. Next, a superconducting NbN film is deposited onto the substrate using direct current magnetron sputtering. The NbN thickness is 5-7 nm, with the sheet resistance $R_s$ =500-600 Ω/sq and the critical temperature $T_c$ =8 K. Details of the NbN deposition process are described in [12]. Then, contact regions are formed with direct photolithography and subsequent reactive ion etching. After the development of the photoresist, very low energy ion oxygen plasma (10-20 eV) is used to remove chemisorbed contaminants and resist residues on superconducting surface.

Next, the resistor body is formed by the photoresist lift-off technique. The molybdenum films are fabricated by DC reactive magnetron sputtering in two different machines, both providing argon ion cleaning-activation before Mo deposition: one at 0.5 keV, the other at 1.5 keV. The thickness of molybdenum films ranges from 20 to 80 nm, which

**Table 1.** Parameters of the studied Type 1 samples.

| Device ID | Energy of Ar ion cleaning, keV | Mo film thikness, nm | Al bandage | Rs of unpatterned Mo film[1], Ω | Rs of Mo measured on sample[2], Ω | Contact resistance $R_0$, Ω | | |
|---|---|---|---|---|---|---|---|---|
| | | | | | | at 300 K | at 20 K | at 2.5 K |
| A-1 | 0.5 | 76.5 | no | 4.4 | 5.2 | 8.1±0.4 | 6.1±0.3 | 5.8±0.3 |
| B-1 | 1.5 | 40 | no | 4.8 | 5.6 | 7.2±0.6 | 5.1±0.3 | 4.4±0.2 |
| A-2 | 0.5 | 70.5 | yes | 4.4 | 5.7 | -0.2±0.3 | --- | -0.2±0.0 |
| B-2 | 1.5 | 40 | yes | 4.8 | 6.0 | -1.4±1.6 | --- | -0.2±0.2 |

[1] measured by Van Der Pau 4-probe method
[2] extracted from device resistance and dimensions of the sample

corresponds to sheet resistances $R_s$ from 19 to 5 Ω/sq (corresponding to resistivities ρ ranging from 20 to 34 μΩ·cm). We adjust deposition time in both machines to have the same sheet resistance.

At the final stage, aluminum outer contact pads for external connection of the chip are made via photolithography lift off technique. The aluminum is deposited by electron beam evaporation. The outer contact pads are formed by optical lithography using positive photoresist S 1805 and e-beam evaporation of 150-nm-thick Al followed by lift-off. The sheet resistance of aluminum is 0.23 Ω/sq, which corresponds to a resistivity of 3.5 μΩ cm. The bandage is fabricated simultaneously with the outer aluminum contacts.

Since the contact resistance of the superconductor/metal interface is greatly influenced by the surface quality, ion beams are used to clean the surface and remove chemically adsorbed contaminants [13]. Thus, the key element of our resistor fabrication process is cleaning-activation of NbN surface in argon plasma with energy either 0.5 or 1.5 keV before deposition of Mo layer in the in-situ process. We also use argon ion-plasma cleaning at 400 eV before Al contact sputtering for better Al to NbN adhesion. For samples with bandage, we also use argon cleaning of Mo surface before Al deposition.

## 3. Results and discussion

### 3.1 Room temperature resistance

Table 1 summarises properties of Type 1 devices. Figure 2(a) shows the dependence of the resistance on the length of the resistor measured at room temperature. The lengths of the resistors are represented by the number of squares $N$ which is determined without considering the contribution of the contact pads. The results are approximated by a linear function $R = R_0 + NR_s$, here $N$ is the number of squares, the straight-line slope coefficient $R_s$ is the sheet resistance of the molybdenum (resistance of a square of the film), and the free term $R_0$ is the contact/interface resistance. As a result of the approximation, the value of $R_s$ is in range 5.2 – 6.0 Ω/sq for all samples. These values are larger than those measured on unpatterned film. The difference can be explained by film oxidation as $R_s$ of unpatterned film is measured right after film deposition, while resistors are measured a couple of weeks later.

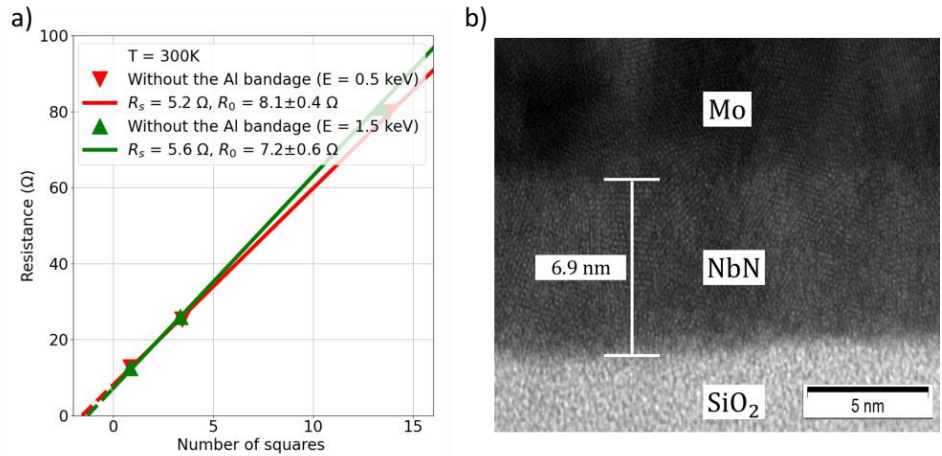

**Figure 2.** (a) Dependence of resistance on the number of squares for resistors without an aluminum layer for the samples with NbN surface cleaned with 0.5 keV argon ions (A-1 sample, red triangles) and 1.5 keV (B-1 sample, green triangles) measured at room temperature; (b) TEM image of SiO2/NbN/Mo interfaces for sample A-1.

**Table 2(a).** XRR analysis of density, thickness, and roughness of the Si/SiO$_2$/NbN layers before deposition of molybdenum film

| Materials | Density, g/cm$^3$ | Thickness, nm | Roughness, nm |
|---|---|---|---|
| Surface layer | 1.2 | 2.1 | 0.5 |
| NbNO$_x$ layer | 4.1 | 2.7 | 0.6 |
| NbN | 7.3 | 6.1 | 0.7 |
| SiO$_2$ | 2.64 | 200 | 0.4 |
| Si (wafer) | 2.32 | - | 0.3 |

**Table 2(b).** XRR analysis of Si/SiO$_2$/NbN/Mo layers after deposition of molybdenum film.

| | XRR after Mo deposition (0.5 keV activation) | | | XRR after Mo deposition (1.5 keV activation) | | |
|---|---|---|---|---|---|---|
| Material | Density, g/cm$^3$ | Thickness, nm | Roughness, nm | Density, g/cm$^3$ | Thickness, nm | Roughness, nm |
| Surface layer | 2.6 | 1.1. | 1.1 | 2.6 | 1.5 | 0.5 |
| Mo | 9.6 | 31.9 | 1.7 | 10.0 | 21.4 | 0.5 |
| NbN | 7.3 | 6.9 | 0.3 | 7.1 | 5.7 | 0.5 |
| SiO$_2$ | 2.64 | 200 | 0.4 | 2.64 | 200 | 0.4 |
| Si (wafer) | 2.32 | - | 0.3 | 2.32 | - | 0.3 |

Both samples A-1 and B-1 (without Al bandage) show additional contact resistance $R_0$ of 7-8 Ω. To understand the origin of excess resistance, we performed TEM analysis of the layers, and X-ray reflectometry (XRR) analysis with Powell method. TEM analysis is performed on a sample fabricated in the same run with the measured device A-1. XRR analysis is performed on an unpatterned film 1 cm x 1 cm in size, fabricated separately by the same route as the measured samples (of course, skipping lithography steps).

The TEM image of the NbN/Mo interface, Fig. 2(b), shows the crystal structure of Mo and NbN without a presence of any additional oxide layers. For this sample NbN surface is activated with 0.5 keV argon ions. XRR data (Table 2a) shows that the 6.1 nm thick NbN film has a 2.7 nm thick layer, presumably NbNO$_x$ oxide, on its top before activation and Mo deposition. After activation of the NbN surface in 0.5 keV argon plasma and deposition of molybdenum, XRR data show a homogeneous 6.9 nm thick NbN layer, which is 0.8 nm thicker than the original NbN layer without any traces of NbNO$_x$. We believe that the NbNO$_x$ layer is almost completely removed in activation process. The remaining thin oxide layer (less than 1 nm) being located between the denser NbN and Mo layers cannot be identified by XRR (Table 2b). Reflectograms of the NbN film before and after activation are shown in the Supplementary in Figure S3. When 1.5 keV ions are used to activate the NbN surface, the NbNO$_x$ layer is completely removed probably together with thin top layer of NbN (Table 2b): NbN thickness is reduced to 5.7 nm (vs original 6.9 nm).

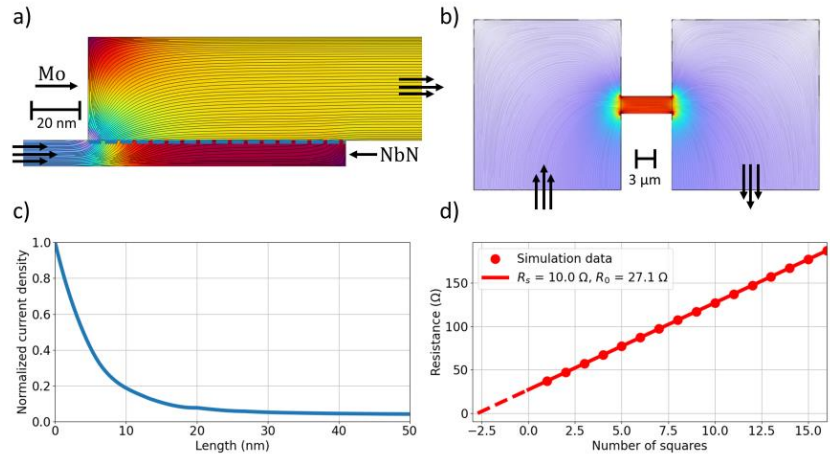

**Figure 3.** Simulation of current distribution in the NbN/Mo interface: cross-section (a) and top view (b). Current density at the NbN/Mo interface along the dashed line of panel (a) shows that almost all the current transfers from NbN to Mo over a length of the order of 10 nm (c). Modeling of the resistance contribution by the contact pads considering that the current is injected from the side of the pad (d), the resistances modeled for various resistor lengths. The contribution of the contact pads is $2.7 R_s$.

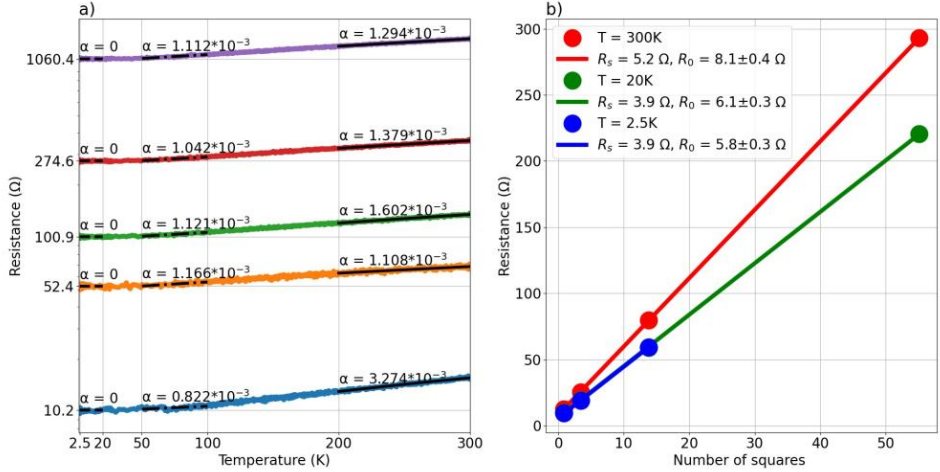

**Figure 4.** (a) Temperature dependences of the resistances for various lengths of resistors. Temperature coefficients α according to approximation formula $(T) = R_{T_0}[1 + \alpha(T - T_0)]$ are given for temperatures 20K, 75K and 300K. (b) Resistances measured for Type 1 sample A-1 at room temperature (red), 20 K (green) and 2.5 K (blue). Change of $R_s$ and $R_0$ when the temperature changes from 300 K to 20 K is due to Mo resistance reduction with temperature, while further change of $R_0$ between 20 K and 2.5 K we attribute to the superconducting transition of NbN under resistor contact pads.

To understand the origin of contact resistance, we performed modeling of the current distribution near the layer boundary. Since at room temperature $R_s$ of NbN is approximately 120 times higher than the $R_s$ of molybdenum, in contact area the current is expected to transfer almost immediately from high-resistivity NbN to low-resistivity Mo as demonstrated by the simulations in Fig. 3(a). Modeling shows that almost all the current transfers into Mo at a distance of about 10 nm from the beginning of the overlap of the layers. Therefore, the resistance of the Mo contact pads contributes to the total resistance. According to $R_0$ obtained from the fit of the plots in Fig. 2(a), excess resistance is 1.3 – 1.5 squares of the film. In our samples current and potential contacts are located on the sides (Fig.1 a,b,d,e). Therefore, we simulated current flow with 90° turn in a contact pad as shown in Fig. 3(b). The simulation made for strips of various lengths shows that two square contact pads contribute $2.7R_s$ ($R_0$ of the fit in Fig. 3(d)) which is larger than experimental value of $R_0$ which can be explained by not exact correspondence between the modelled structure and the real samples.

### 3.2 Resistance at cryogenic temperature

Resistance measurements at cryogenic temperatures were performed using the four-probe method in a Gifford-McMahon cryocooler. Resistance vs temperature is measured for the resistors of different length (Type 1 samples) and is shown in Fig. 4(a). The resistance drops approximately by a factor of 1.3 when cooled down from the room temperature to 50 K and remains practically unchanged with further temperature decrease. For each resistor, its resistance is fitted with formula $R(T) = R_{T_0}[1 + \alpha(T - T_0)]$ in three temperature ranges: 2.5 – 20 K, 20 – 75 K, 75 – 300 K, and coefficient α is extracted.

Figure 4 (b) shows the resistances measured for Type 1 sample A-1 at three temperatures: 300 K, 20 K, and 2.5 K. It is quite natural that in range 300 K – 20 K the contact resistance $R_0$ follows temperature dependence of the Mo film and drops by a factor of 1.3 - 1.4. We believe that a small decrease of $R_0$ from 6.1 to 5.8 Ω (for sample A-1) when the temperature is further reduced to 2.5 K corresponds to superconducting transition of NbN under the contact pads.

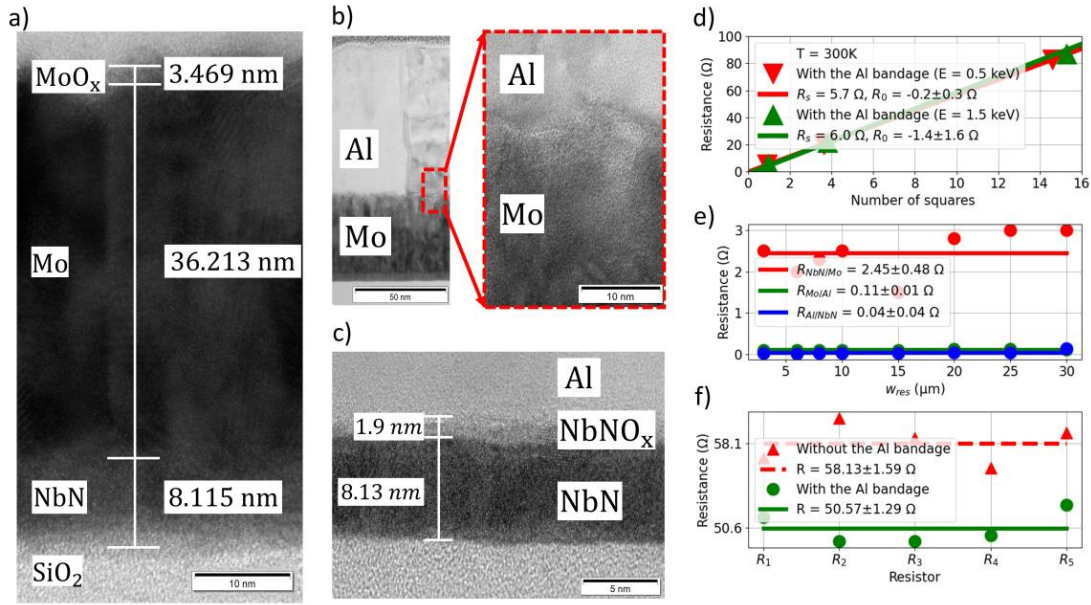

**Figure 5.** (a) NbN and Mo layers before Al deposition, Mo is covered by a $MoO_x$ layer. (b) - (c) TEM images of Mo/Al and NbN/Al interfaces showing oxide layers. (d) Dependence of resistance on the number of squares for resistors with an Al bandage made with activation ion energy of 0.5 keV (red) and 1.5 keV (green). (e) Dependence of interface resistance on the size of the contact pad for each interface NbN/Mo, Mo/Al and Al/NbN measured on Type 3 samples. (f) Spread of resistances for Type 2 samples (five resistors of the same size) fabricated with (green) and without (red) bandage, dashed lines are average resistance over 5 resistors of each sample.

Considering sheet resistance of NbN (about 600 Ω), the residual resistance of 5-6 Ω corresponds to 1/100 of the length of a square of the film. Presumably, at 2.5 K the current flows in NbN contact pad almost up to the end and switches to Mo at the final 1/100 of the length of the pad. A similar behavior is observed for sample B-1 fabricated with 1.5 keV ion activation: $R_0$ is reduced from 7.2 to 5.1 Ω from room temperature to 20 K and drops to 4.4 Ω after transition of NbN in superconducting state at 2.5 K (Table 1).

### 3.3 Interface resistance with Al bandage

A standard way to eliminate contact pads resistance is to use a layer of low-resistance metal such as aluminum. This technique is called "bandage". We applied TEM and XRR to study the quality of Mo and NbN interfaces to aluminum that we use as bandage in our samples. Figure 5 (a-c) presents TEM images of NbN/Mo/$MoO_x$ layers before Al deposition, as well as Mo/Al and NbN/Al layers after Al deposition. Although TEM images visualize $MoO_x$ layer on top of Mo before Al deposition (Fig. 5a), this layer is not observed in XRR analysis after Al deposition, most probably due to the density of molybdenum oxide $MoO_x$ layer which is comparable to that of aluminum. TEM images also show that the NbN/Al boundary has an intermediate layer of $NbNO_x$ oxide with a thickness of about 1.8-1.9 nm (Fig. 5c), which is confirmed by XRR data. Overall comparison of Mo/Al and NbN/Al interfaces studied by XRR is presented in Tables 3(a) and 3(b).

Figure 5(d) presents resistance vs length measured for Type 1 devices with bandage layer (samples A-2 and B-2). Linear approximation shows that contact resistance $R_0$ drops to zero within the accuracy of our measurements (negative $R_0$ produced by the linear fit) without any noticeable difference in activation energies. At 2.5K temperature the contact resistance with the bandage is also less than the accuracy of our measurements for both ion activation energies (Table 1).

Table 3(a). Density, thickness, and roughness of Si/SiO2/NbN/Al from XRR data.

| Materials | Density, g/cm$^3$ | Thickness, nm | Roughness, nm |
|---|---|---|---|
| Al | 2.8 | 140.2 | 2.1 |
| NbNO$_x$ | 6.2 | 1.8 | 0.3 |
| NbN | 7.8 | 6.1 | 0.5 |
| SiO$_2$ | 2.64 | 200 | 0.4 |
| Si (wafer) | 2.32 | - | 0.3 |

Table 3(b). Density, thickness, and roughness of Si/SiO2/NbN/Mo/Al from XRR data.

| Materials | Density, g/cm$^3$ | Thickness, nm | Roughness, nm |
|---|---|---|---|
| Al | 2.7 | 143 | 3.0 |
| Mo | 9.9 | 31.7 | 1.1 |
| NbN | 7.9 | 5.7 | 0.3 |
| SiO$_2$ | 2.64 | 200 | 0.4 |
| Si (wafer) | 2.32 | - | 0.3 |

To ensure reproducible resistor manufacturing, contact pads are typically a few microns in size. Figure 5(e) shows the dependence of the interface resistance measured at room temperature on samples of the third type (Fig.1c) for different areas of the intersecting regions. First, while the resistance of NbN/Mo interface is in range 2 – 3 Ω, the resistances of Mo/Al and NbN/Al are well below 1 Ω (0.11 and 0.04 Ω, respectively). Second, the interface resistance does not depend on the intersection area down to a size of 3 μm × 3 μm. We did not test smaller interface areas as they are less interesting for reproducible fabrication technology, although the simulation results presented in Fig. 3, allow us to hope that the dimensions of the contact pads can be made smaller than 1 μm.

Finally, we tested reproducibility of our technology by fabrication of several identical resistors (Type 2 samples). Figure 5(f) presents resistances measured on two sets of identical samples: with Al bandage (green), and without bandage (red), both measured at 2.5K temperature. The samples are designed for a resistance of 50 Ω. The resistances of samples with a bandage corresponds to the target value of 50 Ω with high accuracy (average measured $R$ is 50.57±1.29 Ω). The average resistance of the samples without a bandage is 58.13 ±1.59 Ω, additional 8 Ω corresponds to the contribution of the contact pads.

## 4. Conclusion

The ion activation of NbN film in argon plasma before Mo resistor deposition improves interface quality by removing NbNO$_x$ layer. There is no noticeable difference between the two studied activation ions energies of 0.5 and 1.5 keV: in both cases we observe additional contact resistance of 7 – 8 Ω at room temperature which we attribute not to the resistance of the interface itself but to the resistance of the Mo contact pad. Our modeling shows that the current transfers from high resistivity NbN to Mo and flows along almost all the length of the Mo contact pad. When cooled down to cryogenic temperatures this contact resistance follows the temperature dependence of Mo resistivity and is reduced to 5 – 6 Ω.

To eliminate the contact pad resistance, we put additional low resistivity aluminum layer ("bandage") on top of the contact pads. In this case the contact resistance is below the accuracy of our measurement which is about 1 Ω. Although the NbN/Al and Mo/Al boundaries feature thin oxide layers it does not affect the interface resistance in all the studied range of contact areas from 30 μm × 30 μm to 3 μm × 3 μm.

We believe that our results are useful for the design of NbN-based superconducting electronics such as photon-number resolving detectors and SNSPD arrays.

## Data availability statement

All data that support the findings of this study are included within the article (and any supplementary files).

## Acknowledgments

This work was supported by the project № 122040800153-0 of Ministry Science and Higher education of the Russian Federation. Fabrication and characterization were carried out at large scale facility complexes for